\documentclass[useAMS,usenatbib]{mn2e}
\usepackage{amssymb,amsmath,amsopn,graphicx,epsfig,subfigure}
\DeclareMathOperator{\argmin}{arg\,min}
\title[Non-linear redundancy calibration]{Non-linear redundancy calibration}
\author[Marthi and Chengalur]{Visweshwar Ram Marthi\thanks{E-mail:
vrmarthi@ncra.tifr.res.in (VRM)} and Jayaram Chengalur\thanks{$\ \ \ \ \ \ \ \ \ \ \ $chengalur@ncra.tifr.res.in (JNC)}\\
\\National Centre for Radio Astrophysics, Tata Institute of Fundamental
Research,\\Post Bag 3, Ganeshkhind, Pune 411007, India
}
\begin{document}

%\date{Accepted 1988 December 15. Received 1988 December 14; in original form 1988 October 11}

\pagerange{\pageref{firstpage}--\pageref{lastpage}} \pubyear{2002}

\maketitle

\label{firstpage}

\begin{abstract}

For radio interferometric arrays with a sufficient number of redundant
spacings the multiplicity of measurements of the same
sky visibility can be used to determine both the antenna gains as well as the
true visibilities. Many of the earlier approaches to this problem focused on linearized
versions of the relation between the measured and the true visibilities.
Here we propose to use a standard non-linear minimization algorithm
to solve for both the antenna gains as well as the true visibilities.
We show through simulations done in the context of the ongoing upgrade
to the Ooty Radio Telescope that the non-linear minimization algorithm is fast
compared to the earlier approaches. Further, unlike the most
straightforward linearized approach, which works with the logarithms of the visibilities and
the gains, the non-linear minimization algorithm leads to unbiased solutions. Finally
we present error estimates for the estimated gains and
visibilities. Monte-Carlo simulations establish that the estimator is
indeed statistically efficient, achieving the Cram$\acute{e}$r-Rao bound.

\end{abstract}


\section{Introduction}
The ``visibilities'' obtained by radio interferometers have to be
corrected for the unknown gains of the antennas along the signal path before they
can be used to image the sky. This is a well known problem (viz.
``calibration'')  and calibration routines  are available in
several different interferometric data reduction packages. A special
case of calibration arises when the array is such that a given
antenna spacing is present multiple times. Each baseline in such
a ``redundant'' baseline set measures the same Fourier component of the sky
brightness distribution. The only differences between the copies
are due to the different multiplicative gains and additive noise
along the different signal paths. For a sufficiently redundant array,
this multiplicity of measurements of the same physical quantity allows 
simultaneous estimation of both the antenna gains as well as the true sky 
visibilities. Some of the earliest work
on ``redundancy calibration'' was done in the context of the
 Westerbork Synthesis Radio Telescope (WSRT) \citep{noor1982,
wier1991a, wier1992}. Interest in redundancy calibration
has been revived in the recent past because a number of new
instruments (e.g. the Low Frequency Array(LOFAR)\citep{falc2006}
and the Murchison Widefield Array(MWA)\citep{lons2009}) have 
some degree of redundancy at certain scales, to exploit specific
engineering advantages. The LOFAR stations, for instance, have a 
redundant $ N \times N$ geometry that allows accurate calibration
of each of the antennas within the station\citep{wijn2012}. 
Further, certain astrophysics and cosmology 
experiments would benefit from redundant configurations for the 
telescope geometry: the cylinder array\citep{pete2006} and 
BAOBAB\citep{pobe2012} are specific examples, while \citet{pars2012a} 
discuss redundant configurations for redshifted $21cm$ studies such 
as the Epoch of Reionization(EoR). 

Many of the algorithms in previous work in redundancy calibration largely used linear least 
squares(LLS) algorithms to solve for the antenna gains and sky visibilities. 
In this paper, we apply a simple non-linear least squares(NLS) 
minimization algorithm and compare its performance against that
of the LLS algorithms. This is done using simulated data, with
the simulations being done in the context of an ongoing upgrade
of the Ooty Radio Telescope.  We also compute error estimates on
the derived solutions, and compare the accuracy of these estimates
against ensemble average errors derived from a Monte-Carlo simulation
and the Cram$\acute{e}$r-Rao bound(CRB).

\section{Redundancy calibration}
\label{sec:redcal}

At any instant of time, a two-element interferometer measures a single 
Fourier mode (called the visibility) of the incident radiation field.  
The (complex) gains along each of the signal paths and the additive
thermal noise cause the measured visibility to differ from 
the true visibility. Mathematically:

\begin{equation}
V_{ij} = g_i g_j^* M_{ij} + N_{ij}
\label{meqn}
\end{equation}

where $g_i$ and $g_j$ are the complex gains associated with the antennas $i$ and $j$ 
respectively, $M_{ij}$ is the true visibility of the sky corresponding to 
the baseline between antennas $i$ and $j$, and $N_{ij}$ is the complex-valued 
additive noise. For the sake of clarity we ignore other factors such as 
the primary beam of the (assumed identical) antennas and polarization leakage which would also cause the 
measured visibility to differ from the true one  \citep[see e.g.][]{tmsbook}.
We also assume that phase variations arising from the ionosphere can 
be lumped together with the complex gain of the receiver chain, and
that the correlator does not introduce any baseline based gains or
errors. We return to a discussion of these assumptions in 
Sec.~\ref{ssec:resultsanddiscuss}.

In order to recover the true visibility, the gains {$g_i$} would have 
to be known, i.e. the interferometer would have to be calibrated. Conceptually,
the simplest way to do this would be to observe a region of the sky
where the true visibilities $M_{ij}$ are known (viz. a {\it calibrator} 
source), and use the observed $V_{ij}$ to solve for the gains. If the 
gain variation timescale is long compared to the time interval between
observations of calibrator sources, the solutions can be interpolated in time to 
obtain the gains at some intermediate time, for e.g. a time when
the {\it target} source, whose true visibilities one would
like to estimate, is being observed. For a multi-element interferometer with $N$ elements,
the number of instantaneous visibility measurements go like $\sim N^2$,
while the number of unknown gains are only $N$. In addition the measured
visibilities must obey the amplitude and phase closure constraints. As is well known, these
facts can be used to iteratively refine both the gains and the
visibilities, (viz. do {\it self} calibration) provided some reasonable initial guess for the gains and the structure 
of the source is available\citep[see e.g.][]{tmsbook}.

Another interesting case arises when the array is redundant, i.e. there
are multiple instances of the same antenna separation. The observed 
visibilities on these redundant baselines would differ only by the
instrumental gains and the additive thermal noise. If there is sufficient 
redundancy this would allow solving for the unknown antenna gains,
as well as the true visibilities(viz. {\it redundancy} calibration),
essentially independent of any assumed model for the sky
\citep[see e.g.][]{corn1999}.  Redundancy calibration algorithms discussed
in the literature are broadly based on linear least-squares methods.
One such example is the LLS algorithm proposed by \citet{noor1982}. We
give a brief description below, and direct the interested reader to
the longer discussion in the paper cited above
and those of \cite{wier1991a, wier1992} and \citet{liu2010}. 

Ignoring the noise term $N_{ij}$ and taking logarithm of equation~(\ref{meqn})
we get:
\begin{equation}
\ln|V_{ij}| = \ln|g_i| + \ln|g_j| + \ln|M_{|i-j|}|
\label{eqn:rca}
\end{equation}
\begin{equation}
\angle V_{ij} = \angle g_i - \angle g_j + \angle M_{|i-j|}
\label{eqn:rcp}
\end{equation} 

The above system of equations can be written in the familiar matrix form 
\begin{equation}
\mathbf{y = A\ x}
\end{equation}
separately for the amplitude and the phase equations, 
where $\mathbf{y}$ is a column vector containing the (amplitudes or phases of
the) observed visibilities on all 
the redundant baselines, $\mathbf{x}$ is a column vector containing the antenna gains 
and the true visibilities on those baselines (with the dimension of $\mathbf{x}$ being 
significantly smaller than the dimension of $\mathbf{y}$) and $\mathbf{A}$ is a sparse matrix 
depending only on the array geometry. For example, for a
uniformly-spaced linear $N$-element array, $\mathbf{y}$ is of length $N(N-1)/2
- 1$ while the 
length of $\mathbf{x}$ is $N + (N-2)$. Specifically, for the amplitude 
equation(~\ref{eqn:rca}): \\
\begin{equation}
\mathbf{y} = \left [ \begin{array}{c} \ln|V_{1,2}| \\ \ln|V_{1,3}| \\ \ln|V_{1,4}|
    \\ \vdots \\ \ln|V_{n-1,n}| \end{array}\right],\hskip 0.1in
\mathbf{x} = \left[ \begin{array}{c} \ln|g_1| \\ \ln|g_2| \\ \ln|g_3| \\ \vdots 
    \\ \ln|g_n| \\ \ln{M_{|1|}} \\ \vdots \\ \ln{M_{|n-2|}} \\ 
  \end{array} \right]
\end{equation}

and\vskip 0.15in

\begin{equation}
\mathbf{A} = \left[ \begin{array}{ccccccccc} 1 & 1 & 0      & \cdots & 1      & 0 &
    \cdots & 0 & 0 \\ 1 & 0 & 1      & \cdots & 0      & 1 & \cdots &
    0 & 0 \\ 1 & 0 & 0      & 1      & \cdots & 0 & \cdots & 0 & 0
    \\ \vdots & & & & \ddots & & & \vdots & \\ 0 & 0 & \cdots & 1 & 1
    & 0 & 1 & 0 & 0 \\ 0 & 0 & \cdots & 1 & 0 & 1 & 0 & 0 & 0 \\ 0 & 0
    & \cdots & 0 & 1 & 1 & 1 & 0 & 0 \\ \end{array} \right] 
\end{equation}

For the phase part of the solution, some elements of the matrix $\mathbf{A}$ undergo 
a sign change corresponding to the complex conjugation of one of the gains, 
without the structure of the matrix itself being altered.  As can be seen, 
the matrix $\mathbf{A}$ is highly sparse, leading to fast computation of the solution
\begin{equation}
\mathbf{x} = \mathbf{A}^{-1}\ \mathbf{y}
\end{equation}
in a single step separately for the amplitudes and the phases. We
clarify that by $\mathbf{A}^{-1}$ we mean
the generalized inverse $(\mathbf{A}^T\mathbf{A})^{-1}\mathbf{A}^T$. Since the matrix
$\mathbf{A}$ is static, its inverse has to be computed only once,
after which it can be used for all time intervals
at which the redundancy calibration needs to be done.

While this method is straightforward to implement, it ignores the additive
noise, and hence is suitable only in high signal-to-noise ratio(SNR) situations.
Liu et al.(2010) discuss the problems that arise if this
algorithm is used in the low SNR regime, and suggest an alternative
algorithm. Their method is based on linearizing the equations
by considering a Taylor series expansion of the complex exponentials 
in equation~(\ref{meqn}).

Following a Taylor series expansion and approximation to the linear
term, the complex gains and visibilities can be rewritten as \vskip 0.1in

\begin{equation}
g_i = g_i^0[1+\Delta\eta_i + i\phi_i]
\end{equation}

\begin{equation}
g_j = g_j^0[1+\Delta\eta_j + i\phi_j]
\end{equation}

\begin{equation}
M_{|i-j|} = M_{|i-j|}^0[1+\Delta\zeta_{|i-j|} + i\theta_{|i-j|}]
\end{equation}

\begin{equation}
V_{ij} = V_{ij}^0[1+\rho_{ij} + i\psi_{ij}]
\end{equation}

\begin{equation}
\delta_{ij} = V_{ij} - V_{ij}^0
\end{equation}

where 
\begin{equation}
V_{ij}^0 = g_i^0g_j^0M_{|i-j|}^0
\end{equation}

The superscript denotes the fiducial guess for the various quantities
that has to be provided at the start of the algorithm. 
Finally, the reduced expression for the correction term reads
as \vskip 0.1in
\begin{equation}
\delta_{ij} = V_{ij}^0[\Delta\eta_i + i\phi_i +
  \Delta\eta_j - i\phi_j + \Delta\zeta_{|i-j|} +
  i\theta_{|i-j|}]
\label{correqn}
\end{equation}

which can be written in matrix form as \vskip 0.1in

\begin{equation}
\mathbf{y = B\ x}
\label{yBx}
\end{equation}

Unlike the earlier case the amplitude and phase equations are coupled. Further, unlike the matrix
$\mathbf{A}$, which depends only on the geometry of the array, the matrix $\mathbf{B}$ depends on
both the geometry of the array and the current estimate of the gains and
true visibilities. It hence has to be updated and a fresh pseudoinverse computed
for each iteration. Note also that one has to supply the fiducial solutions 
$g_i^0$ and $M_{|i-j|}^0$  at the start.  

Both the methods outlined above lend themselves to straightforward 
implementation using any one of the several available linear algebra
libraries. Another recent algorithm of interest is the Weighted
Alternating Least Squares(WALS) algorithm proposed by
\cite{wijn2012}(also see \cite{wijn2010}). The antenna gains and
phases are obtained as the solutions that minimize the covariance matched weighted differences
between the measured and the estimated visibilities:
\begin{equation}
\left\{ \widehat{\mathbf{g}}, \widehat{\mathbf{M}}_0, \widehat{\mathbf{\sigma}}_n\right\}\! =\! \underset{\widehat{\mathbf{g}}, \widehat{\mathbf{M}}_0, \widehat{\mathbf{\sigma}}_n}{\argmin}\|\mathbf{W}_c\!\left(\widehat{\mathbf{V}} - \mathbf{G M}_0
\mathbf{G}^\emph{\small{H}} - \mathbf{\Sigma}_n\right) \mathbf{W}_c\|^2
\label{wals}
\end{equation}

where $\mathbf{\widehat{g}}$ is the vector of the antenna gains to be
estimated and $\mathbf{G}$ is the diagonal matrix of the
gains. $\mathbf{\widehat{M}}_0$ is the Toeplitz matrix of the true 
visibilities(to be estimated) of all baselines obtained from the compact set of
visibilities $\mathbf{M}$ from the unique
baselines. $\mathbf{\widehat{\sigma}}_n$ is the noise power
vector to be estimated and $\mathbf{\Sigma}_n$ is the diagonal
noise covariance matrix. The weighting factor is chosen as
$\mathbf{W}_c = \mathbf{V}^{-1/2}$, where $\mathbf{V}$ is the matrix
of the measured visibilities. A model is assumed initially to estimate the gains $\mathbf{g}$ from
an eigenvalue decomposition. In the next step, these gains are
used to estimate the visibilities $\mathbf{M}_0$. The procedure is
repeated iteratively until a suitable convergence criterion is
satisfied to obtain the estimates $\widehat{\mathbf{g}}$ and
$\widehat{\mathbf{M}}_0$. 

%This is an iterative scheme where the gains are estimated using an eigenvalue
%decomposition. The sky model is iteratively estimated by forcing the
%amplitude and phase constraints on the gains. Unlike the logarithmic
%least squares method which biases the solutions at low SNRs, the WALS
%method is statistically efficient in the sense that it provides
%minimum variance unbiased estimates(MVUEs). We choose not to reproduce the steps of the WALS
%method here merely to preserve brevity. 
%However, we would point out, as and
%when they arise, the many similarities between the WALS method and the
%non-linear least squares minimization algorithm described in the next
%section. 
The first two methods outlined above estimate the amplitudes and
phases of the gains and the visibilities. However the steepest descent
algorithm based on non-linear least squares minimization, described
below, estimates the gains and the visibilities directly as common
numbers on the Argand plane. The phases hence estimated are found to
be free from the errors inherent in the logarithmic method or the
linearized method of \cite{liu2010}, where they are dounf to be
unreliable when large. However alignment of the 
amplitudes and phases would still require external calibration as
explained below. Besides, all the above three methods use matrix
inversion, resulting in $N^4$ operations. However, the WALS method
suggested by \citet{wijn2012} is capable of exploiting the redundant
structure of the problem thereby achieving $N^2$ complexity, putting
it on par with the steepest descent method described below.

\section{steepest descent redundancy calibration}
\label{algorithm}

In the general case of arrays with arbitrary geometry, equation~(\ref{meqn}) is
routinely solved for the unknown antenna gains using non-linear least
squares(NLS) minimization algorithms. It seems reasonable hence to try a
similar method in the redundancy calibration case, with the difference 
being that one would solve not only for the unknown antenna gains, but 
also for the unknown true visibilities.

We begin by defining a real-valued objective function 

\begin{equation}
\Lambda = \sum\limits_i\!\sum\limits_{j > i}\!w_{ij}\|\!\left(V_{ij} -
g_i g_j^* M_{|i-j|}\right)\!\left(V_{ij}^* - g_i^* g_j
M_{|i-j|}^*\right)\!\|
\label{ofn}
\end{equation}

summed over all baselines, where $w_{ij}$ is a real-valued weight. We aim to minimize the objective 
function $\Lambda$ with respect to the complex valued gains $g$ {\it and} 
the true sky visibilities $M$.

%The difference between (\ref{ofn}) and formulation of the WALS
%method (\ref{wals}) is the following: in the WALS method, the additive noise power is
%also a quantity to be estimated. In the non-linear minimization, the
%mean squared error between the measured and the estimated visibilities
%is sought to be matched to the additive noise, similar to the
%chi-squared fitting of a model to a set of measurements. 

 For brevity, henceforth we call the vector 
$\left[\ \mathbf{g}\ \mathbf{M}\ \right]$ the parameter vector 
$\mathbf{\Theta}$.
\begin{displaymath}
\mathbf{\Theta} = \left[\ \mathbf{g}\ \mathbf{M}\ \right] = \left
       [\ g_1,\ g_2,\ \dots\ g_N,\ \ 
M_{|1|},\ M_{|2|},\ \dots\ M_{|L|}\ \right ]
\end{displaymath}

corresponding to a redundant array consisting of $N$ antennas and $L$
redundant baselines. The solutions $\mathbf{\widehat{\Theta}}$, being 
the estimates of $\mathbf{\Theta}$ that minimize $\Lambda$, are
those at which the derivatives of $\Lambda$ with respect to
the elements of $\mathbf{\Theta}$ vanish uniformly, i.e.

\begin{equation}
\frac{\partial\Lambda}{\partial g_k} = 0\ \ \forall\ k \in \{\ 1, 2,
  \dots, N\ \}
\end{equation}

and

\begin{equation}
\frac{\partial\Lambda}{\partial M_{|k-j|}} = 0\ \ \forall\ k\!-\!j \in \{\ 1, 2,
  \dots, L\ \}
\end{equation}

\vskip 0.1in

These equations yield, upon some algebraic manipulation, the solutions

\begin{equation}
g_k = \frac{\sum\limits_{j \neq
    k} w_{kj} g_j M_{|k-j|}^* V_{kj}}{\sum\limits_{j \neq
    k} w_{kj} |g_j|^2\ |M_{|k-j|}|^2}
\label{gainsol}
\end{equation}

and

\begin{equation}
M_{|k-j|} = \frac{\sum\limits_{j > k} g_k^* g_j V_{kj}}{\sum\limits_{j >
    k}w_{kj} |g_k|^2 |g_j|^2}
\label{vissol}
\end{equation}

where the sum is taken over the appropriate redundant baseline sets.

Note that equations (\ref{gainsol}) and (\ref{vissol}) involve 
the (unknown) true gains and visibilities.  This circularity can be
circumvented by taking an iterative approach to the solution: one
takes small steps in the direction of the true solutions, starting from 
an arbitrary initial guess. The corrective steps have to be taken in 
the direction of the negative gradient for the most rapid convergence. 
We redefine quantities in equations (\ref{gainsol}) and (\ref{vissol}) 
as $\mathbf{Q_k}$ and $\mathbf{R_{kj}}$ respectively:

\begin{equation}
\mathbf{Q_k} = \frac{\sum\limits_{j \neq
    k} w_{kj} g_j M_{|k-j|}^* V_{kj}}{\sum\limits_{j \neq
    k} w_{kj} |g_j|^2\ |M_{|k-j|}|^2}
\label{itergainsol}
\end{equation}

\begin{equation}
\mathbf{R_{kj}} = \frac{\sum\limits_{j > k} g_k^* g_j V_{kj}}{\sum\limits_{j >
    k}w_{kj} |g_k|^2 |g_j|^2}
\label{itervissol}
\end{equation}

We additionally define a step size $0 < \alpha < 1$ so that the
solutions can be obtained iteratively as

\begin{equation}
g_k^{n+1} = (1 - \alpha) g_k^n + \alpha \mathbf{Q_k}^n
\label{giter}
\end{equation}

\begin{equation}
M_{|k-j|}^{n+1} = (1 - \alpha) M_{|k-j|}^n + \alpha \mathbf{R_{kj}}^n
\label{miter}
\end{equation}

Of course, the rate of convergence will depend upon the value chosen for
$\alpha$: if it is too small (i.e. $ \alpha <\!< 1$), convergence will be slow, 
on the other hand a large value (i.e. $\alpha \sim 1$), could lead to a 
situations where the algorithm fails to converge. There are well known methods
(such as, for example, the Levenberg-Marquardt method\citep[see e.g.][]{nmrecipes})
for determining the optimal value for $\alpha$: however these require 
computation of the Hessian matrix at each iteration, and are computationally more expensive. 
In our simulations below we use a fixed value of $\alpha$ (i.e. $\sim
0.3$). While a fixed value for $\alpha$ may not be optimal, it more than compensates
for the expenditure of computing the Hessian even if consuming a few more iterations.

We now enumerate the steps involved in steepest descent method for estimating the antenna
gains and the sky visibilities using the non-linear least squares
algorithm given above:
\begin{enumerate}
\item If this is the first time step being solved for, initialize
      the gains and model visibilities to $\mathbf{\Theta} =
      \left[\ \mathbf{g}\ \mathbf{M}\ \right] = \left[  (1, 0),\ (1,
        0),\ (1, 0),\ \ldots(1, 0) \right]$. Otherwise set them to
      the solutions obtained for the last time step. Set the weights $w$ 
      appropriately based on the system temperature. (In the simulation 
      described below we set them all uniformly to unity).  Choose an 
      $\epsilon$ for the convergence criterion (set to $\epsilon = 0.005$
      in our simulations).

\item Integrate  the correlated signal from each antenna pair, i.e. 
      the visibility from each baseline $V_{ij}\ \forall\ i, j \in
      {1,2,\ldots N}, j\!>\!i$,  for the specified time interval.

\item Using the available $\mathbf{\Theta} = \left[\ \mathbf{g}\ \mathbf{M}\
      \right]$ and $w$, compute each $\mathbf{Q_k}$ and $\mathbf{R_{kj}}$ using expressions 
      (\ref{itergainsol}) and (\ref{itervissol}).

\item Update the gains $\mathbf{g}$ and the visibilities $\mathbf{M}$ 
      using equations (\ref{giter}) and (\ref{miter}).

\item Compute the fractional change in each element of $\mathbf{g}$ and 
      $\mathbf{M}$. If the largest fractional change is $> \epsilon$ go to
      step (iii) else stop.
\end{enumerate}

\begin{figure*}
\centering
\includegraphics[scale=0.63]{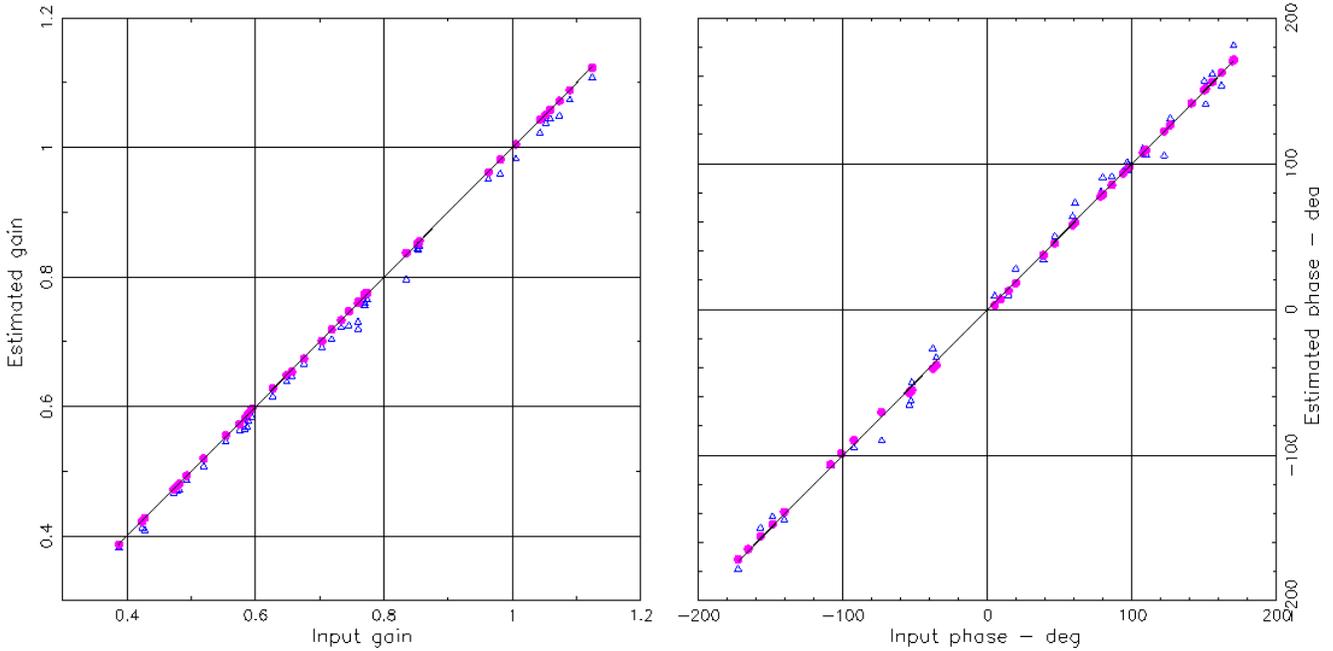}
\caption{The comparison between the antenna gain amplitudes and phases
  as estimated by the logarithmic method of \protect\cite{wier1992} 
  and the  NLS method is described here. The SNR per visibility
  is 10, and the input model is the simplest possible,
  viz. a single source at the phase centre. The solid line represents 
  the ideal situation where the estimated values are equal to the 
  input values. As can be seen, even for this simple model and at 
  this relatively good SNR, the logarithmic method 
  gives biased estimates, shown here as the open circles. In
  contrast, the NLS estimates, shown by the filled triangles, are not biased. 
  \protect\cite{liu2010} discuss a linearized logarithmic method which removes 
  this bias, but which is computationally significantly more 
  expensive.
}
\label{bias-comparison}
\end{figure*}

In the case of routine interferometric calibration, the visibilities $M_{|k-j|}$ 
are known, and hence there is only one equation to work with, viz. (\ref{giter}). 
We point out in passing a further similarity with the calibration of 
non-redundant arrays. Consider the fundamental equation (\ref{meqn}) which 
we rewrite here for the redundant array, ignoring the additive noise term.

\begin{equation}
V_{ij} = g_i g_j^* M_{|i-j|}
\label{meqn_nonoise}
\end{equation}

Scaling $g_i g_j^*$ by some complex constant $z_k$ while simultaneously
scaling $M_{|i-j|}$ by $1/z_k$ leaves the equation unchanged. This is
the well-known ambiguity problem of selfcal\citep{hama2000}. Equivalently,
in the redundant calibration solution, the gain amplitudes are determinable 
only up to an overall scale factor (and the visibility amplitudes to the
inverse square of this scale factor). Similarly the gain phases are determinable
only up to a linear gradient (and the visibility phases to the negative
of this gradient). In this respect, redundancy calibration is
analogous to self-calibration.%, in that in neither can both the total
%source flux and the absolute position of the source be solved for
%simultaneously. 
In the simulations described below, we use the
known input source positions and the known input source fluxes to
determine the overall scale factor and phase gradient. In practice these 
overall scale factors would have to be determined by some external calibration.
Other proposed ways to resolve these ambiguities (see for e.g. the discussion
in  \citet{wijn2012}) is to apply the following constraints to the solutions
\begin{equation}
  \sum\limits_{n=1}^{N}g_n = g_c, \hskip 0.15in \sum\limits_{n=1}^{N}\phi_n = 0, \hskip 0.15in 
  \sum\limits_{n=1}^N \overset{\rightarrow} {\bf r_n}
  \phi_n = 0
\end{equation}

where $g_c$ is some mean gain known apriori, $g_n$ and $\phi_n$ is the
antenna gain amplitude and phase respectively, and $\overset{\rightarrow}{\bf r_n}$ is the position vector of the
$n^{th}$ element in the array. In the case of the ORT, where the telescope is
equatorially mounted, and the baseline lengths do not change with time,
resolving these ambiguities by external calibration and the requirement
that the gain solutions vary slowly but that the visibility solutions
are constant with time is relatively more simple.

\section{Simulations}

\subsection{Description}

The simulations were done in the context of the ongoing upgrade of the
Ooty Radio Telescope (ORT). The ORT is an offset parabolic cylinder with
its long axis parallel to the rotation axis of the earth \citep{swar1971}.
The cylinder is fed by a linear phased dipole array. As part of the upgrade
the signals from these dipoles are digitized and then fed to a correlation
receiver, greatly enhancing the instantaneous field of view and sensitivity
of the telescope. More details of the upgrade and the design of the
new receiver system can be found in \citet{peey2011}.

%runtimes was here but sol.eps comes in here
\begin{figure}
\centering
\includegraphics[scale=0.575]{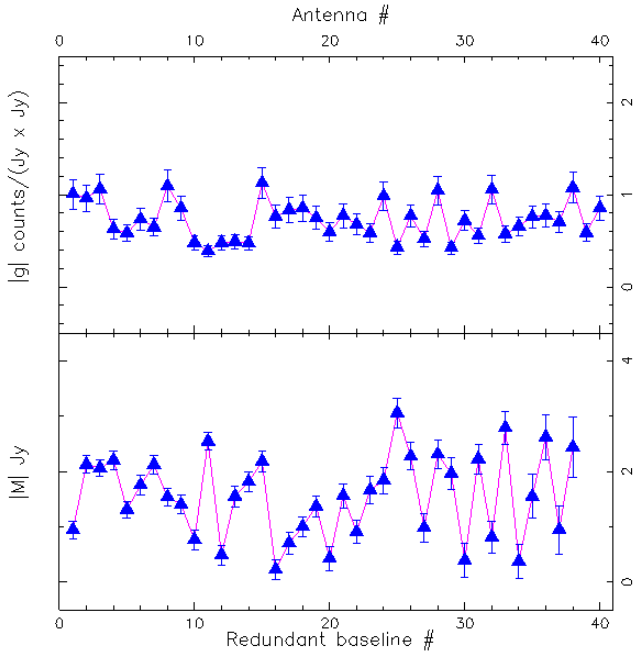}
\caption{The solutions, with their error bars obtained
  from a Monte-Carlo simulation for a realistic sky model, are shown
  here. The signal-to-noise ratio for a $2$~Jy radio source is
  $4$, which can be considered to be the mean SNR over all
  baselines.  The upper panel shows the amplitudes of the gains and
  the lower panel shows the visibilities. The solid line
  connects the true solutions and the filled triangles are the
  estimated solutions. The error bars on the estimated gains have been magnified
  by a factor of $10$ and those on the estimated visibilities by a
  factor of $5$ to enable them to be seen clearly.
}
\label{solbar}
\end{figure}

The redundancy calibration algorithms were checked against simulated data
from the upgraded ORT. The simulator forms a part of a suite of programs 
intended to become an integrated software package for analysis of data 
from the upgraded telescope. The simulator generates the expected
visibilities from a variety of sources including (a) a point source
at the phase centre (b) a collection of point sources randomly distributed
within the field of view and (c) a sky source distribution that follows
the known source counts at the frequency of operation of the telescope,
viz. 325~MHz \citep{wenss}. Since the ORT is oriented in the north-south 
direction and is equatorially mounted, the projected baseline lengths between the feed
dipole elements do not vary when tracking the source. A suitable 
co-ordinate system to work in is hence one in which the $U$ axis is 
oriented east-west, the $V$ axis north-south, and the $W$ axis chosen 
appropriately as to complete a right-handed Cartesian system. Sky positions 
would, as usual, be expressed in terms of the corresponding direction
cosines ($l,m,n$). In such a co-ordinate system, it is easy to show 
that the visibility corresponding to a pair of feed dipoles separated 
by a distance $v = d/\lambda$ for a point source of flux $S$ and 
equatorial co-ordinates ($\alpha_0, \delta_0$) is

\begin{equation}
V(u,v,w) = V(0,v,0) = S. e^{i 2 \pi v m} = S. e^{i 2\pi d \cos\delta_0/\lambda }
\label{ortfourel}
\end{equation}

We note that although the simulations have been done in the specific
context of the ORT, the non-linear steepest descent algorithm is generic
and applicable to any redundant array. In the figure and results presented
below we use a redundant configuration with upto 40 stations (as appropriate for
Phase~I of the ORT upgrade - see \cite{peey2011}). However we have also run the 
simulations for arrays with as few as 5 stations, and find that the NLS 
algorithm works well even in such situations.
For all the simulations Gaussian noise appropriate to the system
temperature, RF bandwidth and integration time was added to the model visibilities.
These simulated visibilities then formed the input to the different
redundancy calibration algorithms.

The matrices $A$ and $B$ in the \cite{wier1992} and \cite{liu2010} 
algorithms are sparse and often singular: techniques like QR
decomposition or singular value decomposition(SVD) would hence have to
be used to solve the corresponding matrix equations. 
In our implementation we have used matrix calls from the GNU Scientific Library(GSL). 
In the non-linear method, as described above, we iterate the equations 
(\ref{giter}) and (\ref{miter}).

\subsection{Results and discussion}
\label{ssec:resultsanddiscuss}

Fig.~\ref{bias-comparison} shows the antenna gains and phases estimated
from a run of the logarithmic redundancy calibration algorithm of
\cite{wier1992}, compared against the same numbers estimated from a run 
of the NLS algorithm, for a signal-to-noise ratio per visibility
of $10$. Although the NLS algorithm works well and
provides unbiased solutions for a realistic sky model, this figure is
shown here purposefully for the simplest input model, i.e. a point source at
the phase centre, so as to illustrate the bias on the gains and
phases at uniform SNRs on all baselines.
It can be seen that despite a reasonable signal-to-noise ratio and a simple
sky model, the logarithmic method gives biased estimates. As described
above, the principal aim of the \cite{liu2010} method was to eliminate
this bias, and in their paper they show results to validate that their
method indeed does avoid the bias. In our implementation too, we find
that the \cite{liu2010} method gives unbiased results; however for
clarity we have shown only results from the NLS method and the \cite{wier1992}
method in Fig.~\ref{bias-comparison}.  From the figure it is clear
that the NLS estimates are unbiased as well. To illustrate further
the quality of the NLS solutions we show in Fig.~\ref{solbar} the 
results from a simulation in which the antenna gains and sky visibilities 
were kept fixed from run to run, but each run had independent noise 
added to the simulated visibilities. The visibilities themselves 
corresponded to a model sky with a source population that matches that 
expected from the source counts at 325~MHz. The figure shows the known 
input parameters to the simulation as well as the mean and rms (computed 
over the different runs, i.e. the ensemble rms) of the  estimated parameters. 
The rms has been scaled by a factor of 10 for the gains and 5 for the
visibilities so that they can be seen clearly in the plot. As can be seen, 
the NLS algorithm does an excellent job of recovering the parameters.

\begin{figure}
\centering
\includegraphics[scale=0.55]{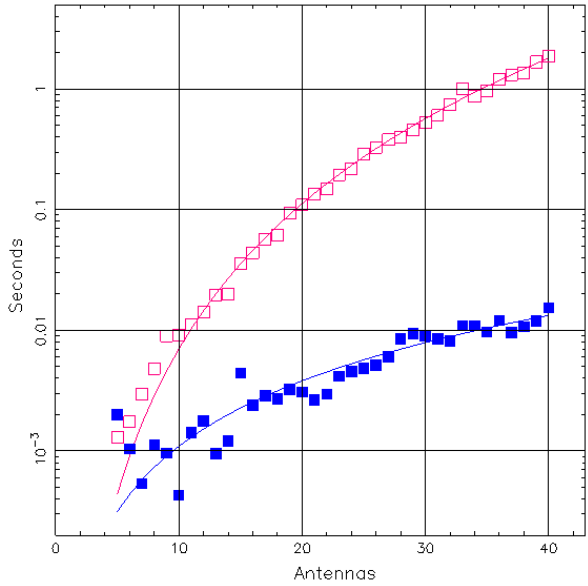}
\caption{Plot showing the comparison between the time taken for the
 two algorithms: the linearized method by \protect\cite{liu2010} and the NLS method
described in this paper. The plot shows the time taken by the algorithms (excluding
the time taken for disk I/O) as a function of the number of antennas
in the array. The sky model is generated from the known source counts
at 325~MHz, and the average SNR per visibility is $\approx$ 4. The
filled squares show the time taken by the NLS algorithm for different number of
antennas, and the open squares show the corresponding time
taken by linearized method of \protect\cite{liu2010}.
The corresponding solid lines show the empirical curves of the form $a.x^n$
with $a = 7.0 \times 10^{-7}, n = 4.0$ 
and $a = 8.2 \times 10^{-6}, n = 2.0$ for the \protect\cite{liu2010} and
NLS methods respectively, clearly reflecting the structure of the algorithm.
}
\label{runtimes}
\end{figure}

Although the linearized method described in \cite{liu2010} avoids
the bias inherent in the logarithmic methods, it is, as described
above, computationally significantly more expensive than the
\cite{wier1992} method. Figure ~\ref{runtimes} shows the time
taken by the \cite{liu2010} and NLS algorithms as a function of
the total number of antennas. The run times exclude the time taken
for data I/O from the disk. The linearized algorithm of
\cite{liu2010} clearly shows a $N^4$ dependence, whereas the
NLS\footnote{The referee of this paper also
communicated to us that the NEWSTAR package also uses a steepest 
descent method which leads to $N^2$ performance when the structure of
the problem is properly exploited.}
algorithm behaves as $N^2$. Note that there was no explicit 
multi-threading in our implementation, and no compiler optimization
was used. The NLS algorithm is significantly faster and is very
effective in real-time calibration.

Since calibration is essentially an estimation problem, it is natural
to ask what the errors on the estimated parameters (i.e. the gains
and visibilities in this case) are, and how the errors obtained via
the NLS algorithm compare with fundamental bounds on the error, viz.
the CRB. As is well known \citep[see e.g.][]{nmrecipes} the covariance 
matrix $\mathbf{\Sigma_{\vartheta}}$ of the estimated parameters is related 
to the Hessian matrix $\mathbf{\mathcal{H}}$ as

\begin{equation}
\mathbf{\mathcal{H}\left(\vartheta\right)} = \mathbf{\Sigma_\vartheta^{-1}}
\end{equation}

\begin{figure}
\centering
\includegraphics[scale=0.55]{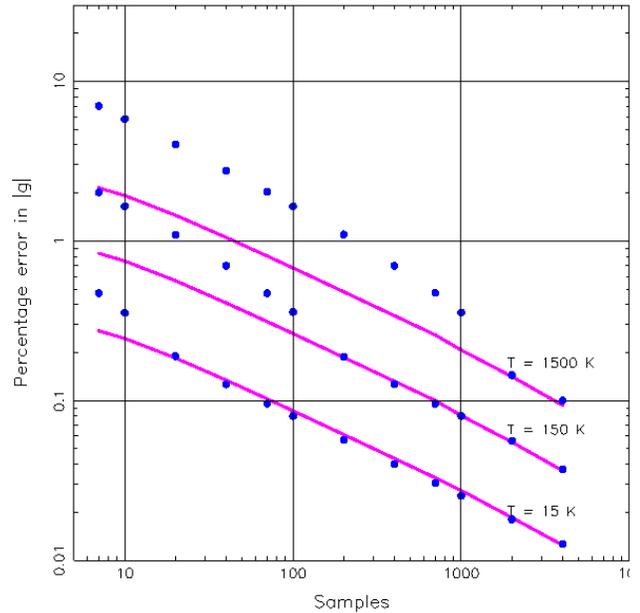}
\caption{This figure shows the behaviour of the error on the
  antenna-averaged gains for three different system temperatures. The filled circles
  represent ensemble errors obtained from the Monte Carlo run, whereas
  the solid line is the Cram$\acute{e}$r-Rao bound. At lower system
  temperatures, the errors reach the CRB upon integration of fewer
  samples. At $T_{sys}$ = 1500 K, for example, many more
  samples would have to integrated than at $T_{sys}$ = 150 K to attain
  the CRB.
}
\label{crbstats}
\end{figure}

Further, if the measurement errors are independent and identically
distributed (i.i.d) Gaussian noise, then the Hessian matrix is equal to 
the Fisher information matrix and the variances estimated from 
the Hessian matrix correspond to the Cram$\acute{e}$r-Rao
bound\citep[see e.g.][]{poor1994}. The parameter errors estimated
in this way would hence be a lower bound to the true error.
Fig.~\ref{crbstats} shows a comparison between the mean ensemble error
obtained from the simulations for runs with different system temperatures
(i.e. signal-to-noise ratio). The ensemble errors attain the CRB when
the SNR $\gtrsim 10$.

\begin{figure*}
\centering
\begin{minipage}{170mm}
\subfigure[Hessian matrix of the estimated parameters
]{\label{subfig:Hessian}\includegraphics[scale=0.48]{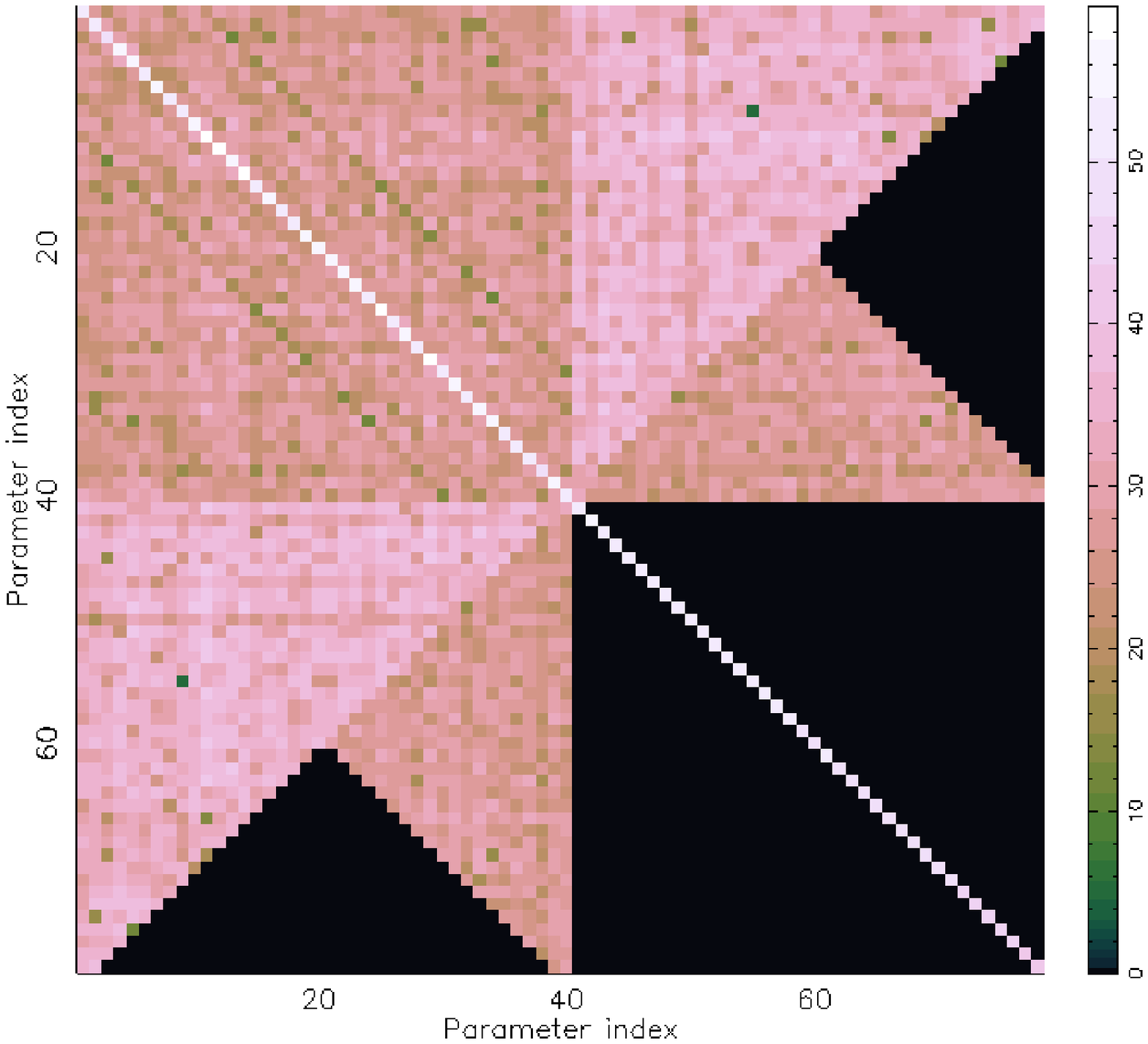}}
\hskip 10mm
\subfigure[Covariance matrix of the estimated
  parameters]{\label{subfig:covar}\includegraphics[scale=0.48]{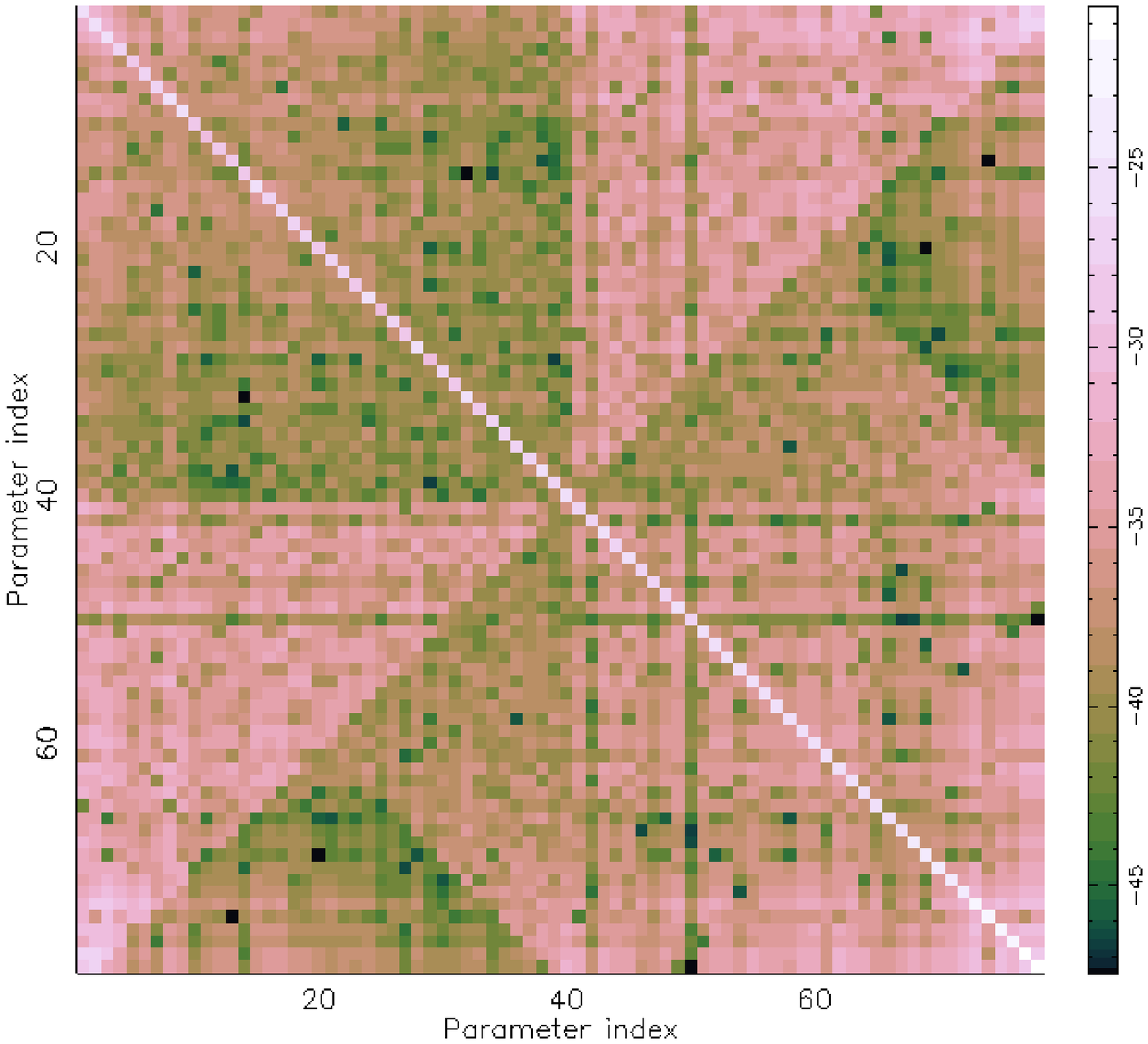}}
\end{minipage}
\caption{The kite plots: \subref{subfig:Hessian} shows the magnitude of 
  the complex elements of the Hermitian-symmetric Hessian matrix, and 
  \subref{subfig:covar} shows the magnitude of the complex elements of 
  the Hermitian-symmetric covariance matrix, both given in
  decibel($10\ log_{10}(.)$) scale to accommodate the large contrast. The matrices 
  are both $78 \times 78$ elements across, with the first $40$ elements 
  on each side corresponding to the errors on estimated complex antenna gains 
  $\widehat{g}_i$, and the next $38$ elements, the errors on the estimated model 
  visibilities $\widehat{M}_{|i-j|}$. The diagonal elements of the
  matrices are, of course, real. The full range of grayscale intensity
  is exploited here using the ``cubehelix'' mapping scheme\citep{dag2011}.
}
\label{errormatrix}
\end{figure*}

Finally, in Fig. \ref{errormatrix} we show the ``kite plots'':
the Hessian and covariance matrices. These matrices
are clearly diagonally dominant. A good approximation to the variance can
hence be obtained quickly by approximating the Hessian matrix to be diagonal.
In the simulation described above, this approximation leads to a difference
of only $\sim 1\%$ in the values of the estimated standard deviation.
Since the diagonal approximation simplifies the inversion of the Hessian,
we could adopt this approximation in the Levenberg-Marquardt algorithm to refine the
step size after each iteration. However, since we find that a constant
step-size is adequate we do not pursue this any further here.
% figure here

  As stated in Sec.~\ref{sec:redcal} we have been assuming that the 
  beamshapes of the individual antenna elements are identical, and that
  the ionospheric phases can be lumped together with the electronic gains
  of the elements. In general these assumptions are only approximately true.
  Further, we have been only dealing with a scalar equation, whereas,
  in the presence of phenomena which mix the two polarizations (e.g.
  differential faraday rotation across the array, receptors whose
  response is not perfectly orthogonal, (i.e. suffer from ``leakage'',
  ``ellipticity'', which themselves could in general vary across the
  field of view) it is not possible to decouple the calibration of 
  the nominally orthogonal polarizations. High dynamic range imaging would
  require one to address all of these issues, and there have been a number
  of algorithms proposed (dubbed ``3GC'' calibration
  algorithms. See for e.g. \cite{noor2010, smi2011b, bhat2008}) to address these 
  issues. A further issue that is relevant in dealing with dipole arrays 
  (as is the case of the focal array at the  ORT) is that there could 
  be mutual coupling between the elements. Addressing these issues for 
  the ORT is beyond the scope of the current paper, but  we plan to 
  return to these in future work. We discuss below some  considerations 
  which suggest that the approximations that we have made are reasonable
  to first order for the problem at hand. Firstly, the ORT measures
  only the polarization along the length of the telescope, i.e. north-south. 
  The dipoles are hence arranged end-to-end, a configuration that one would
  intuitively expect to minimize mutual coupling. Indeed measurements of the
  sensitivity indicate that the sensitivity of the ORT increases linearly
  with the number of dipole signals added, again suggesting that the mutual
  coupling can be ignored to first order. Further since the ORT is equatorially
  mounted, the beams do not rotate in the sky as it tracks the source, and
  hence to the extent that the different sections have identical beams, 
  the calibration issues are greatly simplified. Interestingly, this means
  that the baseline lengths will also be fixed as it tracks the source, 
  meaning that each element pair measures the same sky visibility at all 
  times. Finally since the ``array'' is small (530m), Phase I of the upgrade
  (which breaks the array up into 40 elements) falls within Lonsdale's 
  regime 1\citep{lons2005}, where the traditional solution
  to the ionospheric phase suffices. Phase II, where every 2m segment
  of the telescope will be digitized (i.e. a compact array, wide field of
  view, although note that even in this case, the E-W field of view is
  limited to $2^o$ because of the reflector) would fall into Lonsdale's 
  case 3 where each source in the field of view could be shifted by a 
  time variable offset, that would need to be
  calibrated for. We note that the NLS algorithm suggested above would also
  be useful to quantify the importance of  effects. Residuals (obtained
  after calibration as described here) between the visibilities measured 
  on nominally redundant baselines would be indicative of the importance
  of the direction dependent effects.

%sol.eps was here but runtimes.eps comes in here

\section{Conclusions}

We apply a standard iterative non-linear least squares (NLS) 
minimization algorithm (i.e. steepest descent) to the problem 
of redundancy  calibration. We apply this method to simulated 
data; the simulations were done in the context of the ongoing upgrade of
the ORT. We find that the NLS algorithm is fast and accurate 
compared to the linear least squares methods that has been used 
in the past. The linear method works with the logarithm of 
the antenna gains and the sky visibilities, and hence leads 
to biased estimates of these parameters even at modest signal-to-noise
ratios. Linearization using a Taylor expansion of 
the logarithm in the equations used in the linear method has
been proposed to circumvent this problem. In comparison to this,
we find that the NLS algorithm is fast as well as simple
to implement. We also investigate the accuracy of the estimates 
obtained by the NLS algorithm and find from a Monte-Carlo 
simulation that the ensemble parameter errors attain the
Cram$\acute{e}$r-Rao bound at moderate signal-to-noise ratios.

\section{Acknowledgments}
The authors wish to thank Mihir Arjunwadkar(NCRA/CMS), Shiv Sethi(RRI) and
Rajaram Nityananda(NCRA) for many useful discussions and inputs. They
also thank C R Subrahmanya(RRI), Peeyush Prasad(ASTRON) and P K
Manoharan(RAC) for their help during the course
of this work. We owe our many thanks to the engineering support staff at
RAC, Ooty for their contribution to the design and installation of the
receiver. The authors wish to thank Jan Noordam for his comments which
have helped improve this paper significantly.

\label{lastpage}

\end{document}